\documentclass[runningheads,a4paper,oribibl]{llncs}

\usepackage{amssymb}
\usepackage{graphicx}
\usepackage[table]{xcolor}
\usepackage[font=small,labelfont=bf]{caption}
\usepackage{graphicx}
\usepackage{caption}
\usepackage{amsmath}

\usepackage{url}
\usepackage{float}

\newcommand{\ignore}[1]{}

\usepackage{url}

\urldef{\mailsa}\path|imrul@mail.usf.edu, mithuns@sp.com,jacob@ua.edu|

\begin{document}




%
%
%


\title{Product Backlog Rating: A Case Study On Measuring Test Quality In Scrum}
\titlerunning{A Case Study On Measuring Test Quality In Scrum}

\author{Imrul Kayes\inst{1}\thanks{Most of the work was done when the author was a quality assurance engineer at SoftwarePeople.}, 
Mithun Sarker\inst{2}
\and Jacob Chakareski \inst{3}\\
}
\institute{Computer Science \& Engineering \\University of South Florida, Tampa, FL, USA \and
SoftwarePeople\\
Copenhagen, Denmark \and
Electrical and Computer Engineering\\
University of Alabama, Tuscaloosa, Alabama, USA
\mailsa}

\maketitle

%
%


\begin{abstract}

Agile software development methodologies focus on software projects which are behind schedule or highly likely to have a problematic development phase. 
In the last decade, Agile methods have transformed from cult techniques to mainstream methodologies.
Scrum, an Agile software development method, has been widely adopted due to its adaptive nature.

This paper presents a metric that measures the quality of the testing process in a Scrum process. 
As product quality and process quality correlate, improved test quality can ensure high quality products. 
Also, gaining experience from eight years of successful Scrum implementation at SoftwarePeople, we describe the Scrum process emphasizing the testing process.
We propose a metric Product Backlog Rating (PBR) to assess the testing process in Scrum.
PBR considers the complexity of the features to be developed in an iteration of Scrum, assesses test ratings and offers a numerical score of the testing process. 
This metric is able to provide a comprehensive overview of the testing process over the development cycle of a product. 

We present a case study which shows how the metric is used at SoftwarePeople.
The case study explains some features that have been developed in a Sprint in terms of feature complexity and potential test assessment difficulties and shows how PBR is calculated during the Sprint.
We propose a test process assessment metric that provides insights into the Scrum testing process.
However, the metric needs further evaluation considering associated resources (e.g., quality assurance engineers, the length of the Scrum cycle).
\end{abstract}







\section{Introduction}

Agile software development is a group of software development methods, which attempts to provide an answer to the eager software business community asking for lightweight along with faster and nimbler software development processes~\cite{abrahamsson2002agile}.  
The success of an Agile process depends on the development team's ability to facilitate continuous improvement of the end product in order to reduce cost, effort, and time-to-market, but at the same time to restrain the ever-increasing complexity and size of software systems~\cite{rech2007handling}.
Agile methods focus on close communication and fast feedback to foster mutual respect and cooperation between stakeholders and developers~\cite{churchville2008agile}. 

Although there has been a growing debate on how Agile methods could be practiced in software development, Agile software development methodologies are widely accepted.
According to the 2012 CHAOS report from the Standish Group, Agile projects are successful three times more often than non-agile projects~\cite{Chan2013Agile}.
This statistic is consistent with the Standish Group's 2006 report where it stated that 41\% of Agile projects succeeded as opposed to 16\% of waterfall projects~\cite{Wailgum2007Agility}. 
Moreover, in the 2012 report, 49\% of the stakeholders say that most of their companies are using Agile development.
The report also states that the number of those software development organizations who plan to implement Agile development in future projects has increased from 59\% in 2011 to 83\% in 2012.
Research shows that adopting Agile methods improves the overall management of the development process and customer relationships~\cite{ceschi2005plan}, decreases the amount of overtime and increases customer satisfaction~\cite{mann2005scrum}. 


Agile development is not a methodology in itself, rather  it is an umbrella term that describes several iterative and incremental software development methodologies.
The most popular Agile methodologies include Extreme Programming (XP), Scrum, Crystal, Dynamic Systems Development Method (DSDM), Feature-Driven Development (FDD), Lean Development and so on.
However, Scrum has long been recognized as the most popular Agile method.
According to~\cite{schwaber2007agile,schwaber2004agile}, the most commonly used Agile method is Scrum. 
Standish Group's recent report~\cite{Chan2013Agile} acknowledges this fact and states that Scrum is the most adopted (52\%) Agile methodology.
More than $14500$ Scrum Master certificates were issued since 2003~\cite{alliance2007scrum} and in the last few years several successful implementations of Scrum have been reported in the literature~\cite{schatz2005agile,upender2005agile,mahnic2005agile,mann2005scrum}.

Scrum is an iterative and incremental Agile software development method.
It is driven by a Product Backlog, which contains all requirements of the product. 
A Scrum process begins by reviewing a Product Backlog with the product owner. 
Highest priority features are identified and prioritized and a Scrum team estimates how many will fit into a Sprint. 
These features are then split into Sprint Backlog items. 
A Sprint is a predefined period of time, usually $2$ to $4$ weeks.
During this time, a Scrum team analyzes, designs, constructs, tests and documents the selected features~\cite{schwaber2004agile,smith2009becoming}.

An integral part of the Scrum process is testing the product.
Worldwide adoption of the Scrum method has motivated research of  testing in Scrum. 
Quality has a strong subjective element and as a concept is difficult to define and describe~\cite{bratthall2000understanding}.
Quality experts have  proposed software quality models (e.g.,~\cite{budgen2003software,crosby1979quality,galin2004software}), but those are not supposed to measure the quality of the quality endurance process but to measure representative attributes such that when combined can provide some notion of quality of the  product.

In this paper, we focus on measuring the quality of the testing process in a Scrum process.
The relationship between product quality and process quality is complex.
First, it is challenging to measure software quality attributes, even after using the software for a long period.
Second, it is difficult to articulate how process characteristics influence these attributes. 
However, pragmatic experience has shown that process quality has a significant influence on the quality of the software~\cite{fuggetta2000software}. 
Moreover, process quality improvement results in fewer bugs in the delivered software~\cite{sommerville1998book}. 
Hence, a fundamental assumption of quality management is that the quality of the development and the testing processes directly affect the quality of the delivered product~\cite{cugola1998software,fuggetta2000software,kitchenham1996software}. 
In this paper, we do not attempt to measure the product quality in Scrum, rather we are interested in assessing the quality of the testing process in Scrum.
To the best our knowledge, we are the first to evaluate the quality of a software testing process in an Agile process.

This paper presents a metric, Product Backlog Rating (PBR), which measures the quality of the testing process in Scrum.
PBR considers complexity of the features to be developed in a Sprint, assess test ratings and quantizes quality of the testing process.
Using this metric test quality in a Sprint can be measured. 
PBR scores from Sprints could be used to temporally evaluate the testing process.
Furthermore, if an organization has multiple Scrum teams, it can compare testing processes among those teams.
This comparison could help an organization  to get an overview of the overall testing process in the organization.  

The contributions of this work are as follows:
\begin{itemize}
  \item We first present how the Scrum method is practiced at SoftwarePeople\footnote{http://www.enfatico.com/}.
  SoftwarePeople is a Denmark-based Marketing Operation Management (MOM) solution provider, which adopted the Scrum process as a product development method about eight years ago and is successfully practicing the process.
  We explain the core elements of the Scrum process such as Artifacts, Roles and Events based on our Scrum implementation at SoftwarePeople.
  While explaining those core elements, we emphasize the role of a quality person in a Scrum process. 
 We also present the testing process in Scrum. 

  \item We introduce a metric Product Backlog Rating (PBR) to measure the testing phase in Scrum.
  PBR considers Product Complexity Level (PCL) and Test Assessment Rating (TAR) and offers a numerical score.
  We explain a wide range of factors; which could be leveraged to evaluate Product Complexity Level and Test Assessment Rating.
  
  \item We show the applicability of PBR by presenting a case study from SoftwarePeople.
  The case study picks features that were developed in a Sprint and shows how PBR was calculated during the Sprint.
  
  \end{itemize}
  
The rest of the paper is organized as follows. 
Section~\ref{scrum} explains the Scrum method emphasizing software testing in Scrum. 
Section~\ref{PBR} introduces the PBR metric. 
We present PCL and TAR factors  in Section~\ref{pcl} and Section~\ref{tar} respectively. 
Section~\ref{casestudy} presents a case study. 
Section~\ref{related} reviews related work and Section~\ref{conclusion} concludes. 

\section{The Scrum Process}
\label{scrum}

Scrum is an iterative and incremental Agile software development framework for managing complex product development.
The core characteristics of a Scrum process are--- it is lightweight, simple to understand, but difficult to master~\cite{Schwaber2013Scrum}. 
The term ``Scrum'' originally came from a popular sport Rugby, in which  fifteen players on each team compete against each other~\cite{cho2010exploratory}. 
While in Rugby the term Scrum refers to the strategy used for getting an out-of-play ball back into play, the term ``Scrum''  in product development was first coined by Hirotaka Takeuchi and Ikujiro Nonaka back in 1986~\cite{takeuchi1986new}.
Interestingly, this term was not directly targeted to the software development.
Inspired by some case studies from manufacturing firms such as automotive, photocopier and printer industries, Takeuchi and Nonaka envisioned a new approach to commercial product development that would increase speed and flexibility as opposed to a traditional, sequential approach.
In 1993, at Easel Corporation, Jeff Sutherland first applied the Scrum process to software development teams when they built the first object-oriented design and analysis (OOAD) tool that involved round-trip engineering~\cite{sutherland2004agile}.

The process of Scrum consists of Artifacts, Roles, and Events~\cite{Schwaber2013Scrum}.
While the core attributes of a Scrum process are well established, organizations have adopted different implementations in practice (e.g., \cite{Gannon2013Scrum, upender2005agile,schatz2005agile,Srinivasan2009Scrum,Smits2007Scrum}).
We will discuss the components of the Scrum process and provide a detailed workflow from a quality perspective, gaining experience from eight years of successful Scrum implementation and practices at SoftwarePeople.
We do not argue that our implementation is perfect (as Scrum is simple to understand but difficult to master~\cite{Schwaber2013Scrum}). However, we believe that SoftwarePeople's Scrum practices could shed more lights on both development and testing.

\subsection{Scrum Artifacts}
Scrum has three artifacts: Product Backlog, Sprint Backlog and Burndown Chart~\cite{Schwaber2013Scrum}.
At SoftwarePeople, we use an additional artifact---Dashboard.

\emph{Product Backlog (PBL):} A Product Backlog (PBL) is a prioritized list of requirements of the product. 
The log contains customer requirements (including functional and non-functional), as well as technical requirements. 
The Product owner and other stakeholders write the Product Backlog as a form of user stories.
SoftwarePeople uses a web-based service to maintain the Product Backlog.
The Product owner and other stakeholders add user stories and prioritize user stories using personalized user interfaces.
Other members of the Scrum team have access to the service, but only the Product owner and stakeholders have privileges to edit (a history of the edition is also maintained).
Easy to access and transparent Product Backlog help to reduce manual documentation and management overhead, a practice suggested by the Agile development~\cite{shore2008art}. 

\emph{Sprint Backlog}: The Sprint Backlog is the set of items from the Product Backlog that the team must address during the next Sprint.
More specifically, the most prioritized requirements  from the Product Backlog are chosen and split into Sprint Backlog items.

\emph{Burndown Chart}: A burndown chart shows the daily progress of a Sprint over the Sprint's length.
The Burndown Chart is an indicator of  ``how much'' work has been done and ``how far'' to go.

\emph{Dashboard}: A Dashboard is a web-based service, which represents the status of a Sprint Backlog item.
Sprint Backlog items are divided into smaller pieces (tasks) in a Sprint planning meeting.
There are five statuses of a task: ``not done'', ``in progress'', ``in review'', ``quality assurance'' and ``done''.
At the beginning of a Sprint, all  tasks are uploaded in the Dashboard and their statues are set as ``not done''.
Developers pick the tasks from the Dashboard by moving the tasks from ``not done'' to ``in progress''.
Once the development is done, a developer places a task to ``in review'' list.
Another developer who is not involved in writing the code of the task reviews the task and places it to ``quality assurance'' list.
A quality person (e.g., tester or quality assurance engineer) tests and sets a task's status as ``done''.
However, if the task is not testable (e.g., a data adapter to fetch data), a developer directly sets the status of the task from ``in review'' to ``done''. 

\subsection{Scrum Roles}
There are four roles in a Scrum team: a Product owner, a Scrum Master, Developers and a QA (quality assurance) engineer (again note that our adoption of the Scrum process might differ from others---e.g., in~\cite{Gannon2013Scrum} the author mentioned three roles).

\emph{Product owner:} The Product owner manages the Product Backlog.
A Product owner includes stories of user requirements, prioritizes the Product Backlog and ensures that the Product Backlog is accessible, transparent, and clear to all other members.
The Product owner directs the Scrum team toward ``what to do next''.

\emph{Scrum master:} A Scrum master facilitates the Scrum and helps the development team to create high-value products by ensuring that the Scrum team adheres to the Scrum theory, practices, and rules.
He removes all the impediments and outside distracting influences that come along in the process of achieving Sprint goals.
A Scrum master also co-operates with the Product owner in terms of Product Backlog management and prioritization of Product Backlog items to maximize value.

\emph{Developers:} Developers  create and deliver a potentially releasable increment of ``Done'' product at the end of each Sprint. 
Developers may have specialized skills and areas of focus, but the team as a whole is cross-functional in a sense that it possesses all   skills a team required to create a product increment.
While the exact numbers of the developers for a successful Scrum implementation is debatable (a typical Scrum recommendation is less than ten developers~\cite{rising2000scrum}), SoftwarePeople Scrum teams consist of four to six developers per team.

\emph{Quality Assurance (QA) engineer:} The focus of a Quality Assurance (QA) engineer  in  a Scrum team is to ensure the quality of the deliverables. 
Contrary to the synchronous developments of a traditional waterfall project, Scrum expects asynchronous development activities (e.g., user requirements could be changed a lot during the development cycle).
In waterfall developments, a QA role is only involved at the very end of the project.
But in Scrum, the role is activated from the beginning of a Sprint.
The QA role is not merely involved in writing and executing test cases (e.g., waterfall developments), rather the role is more motivated toward ensuring product quality.
In a software development process, requirement and design errors are more expensive to correct-typically, about 100 times
more expensive than implementation errors~\cite{boehmsoftware,Ramamoorthysoftware}.
Also bug dependencies might worsen the situation~\cite{Kayes2011Fault}.
So, Scrum incorporates quality thinking via a QA role, from the beginning of a Sprint.

In the Sprint planning meeting, a QA engineer helps the team to uncover complex business logic and thus enables to project a correct estimate of the stories.
A QA engineer can identify complex and negative test case scenarios which help the team to discover what sub-components of the product could be affected due to the Sprint.
This early determination could significantly reduce bugs when an iteration is done.

During the initial phase of a Sprint, when not much testable Sprint Backlog items are done, a QA engineer writes system test cases (manual and automatic), prepares test data, clarifies the requirements (if something is still unclear) with the Product Owner and conveys requirement updates to the development team and ensures the test environment.

During the Sprint, a QA engineer writes ``checklist'' for Sprint Backlog items.
Checklists are concise versions of a full test case.
After developing a Sprint Backlog item, developers execute the checklist to ensure that all required functional works are done.
Checklists help determining early bugs, even before a QA engineer starts any testing, which significantly reduces bugs.
Moreover, a QA engineer helps developers to write unit tests and he ensures that code reviews are done in a timely manner.

A QA engineer also shows a ``Sneak Peek'' of the product to the Product owner during the halfway of the Sprint.
A Sneak Peek is a demonstration of what has been done until that period.
The Product owner might change some requirements after getting an initial feel on how the requirements will take shape in the long run.
By enabling early determination of requirement changes, Sneak Peek helps to improve the quality of the product.
And finally, a QA engineer is responsible for testing of the developed features based on his test cases (in our case the tasks who have done  statues on the Dashboard).
We will not elaborate the testing, as the testing tasks are similar to other software development methodologies (for details of testing activities, we refer readers~\cite{beizer1995black,perry1999software,majchrzak2012improving}).

\begin{figure*}[htbp]
\centering
\includegraphics[height=5.90cm]{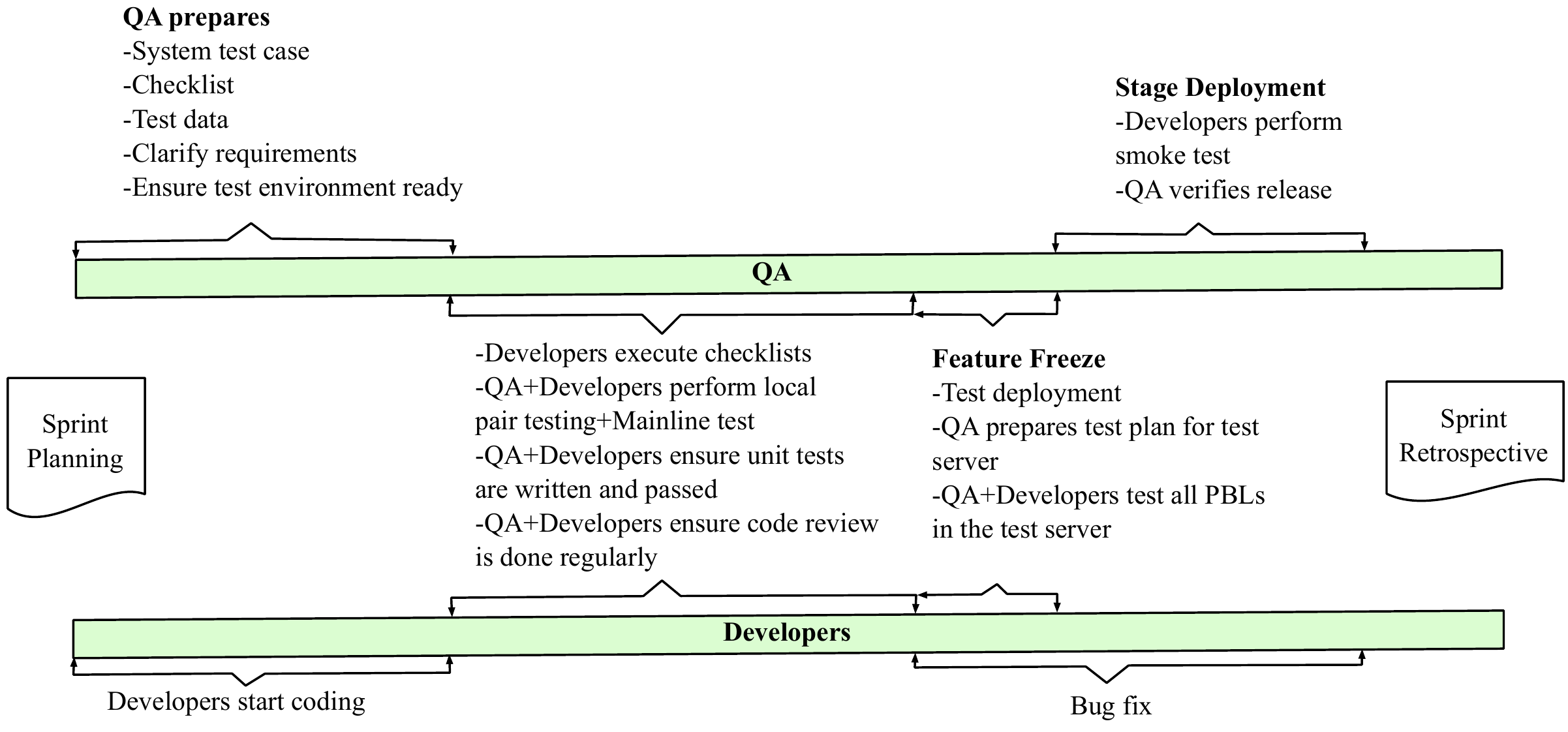}
\caption{Testing activities in a Scrum process.}
\label{fig:QA_in_scrum}
\end{figure*}

When all deliverables of a Sprint are developed, they are deployed to a test server for Regression testing.
The QA engineer prepares a test plan and both developers and the QA engineer run the test cases in the test server.
Once the Regression testing is done, deliverables are deployed to another server named ``Stage''.
On stage server, developers perform Smoke tests and the QA engineer verifies features of the Scrum.
User acceptance testers (not a part of the scrum team) also conduct acceptance tests on the stage server.
Finally, when needed, a Technical team (also not a part of the Scrum team) deploys the deliverables from the Stage server to a Production server. 
Figure~\ref{fig:QA_in_scrum} shows testing activities in SoftwarePeople's Scrum and how developers are also involved in testing along with the QA engineer.

\subsection{Scrum Events}

\emph{Sprint planning meeting:} The purpose of a Sprint planning meeting is to determine what will be done in the Sprint, and how that work will get done~\cite{Schwaber2013Scrum}.
The Sprint Backlog items are first created based on the Product Backlog items to be completed in the Sprint. 
Sprint Backlog items are broken down into units, called tasks.
Each task is timeboxed in a sense that it is possible to develop that task within a half day (three hours) or a full day (six hours).

\emph{Daily Scrum:} A Daily Scrum is a time-boxed (usually fifteen minutes) stand up meeting held every day before the Daily Sprint.
The Scrum master, Developers and the QA engineer participate in the meeting and answer what they have accomplished in the previous day, what they want to commit to do on that day and whether there are impediments.
Impediments (e.g., task dependencies) are resolved immediately if possible, otherwise the Scrum master attempts for a resolution as soon as possible.

\emph{Sprint Review:} There is a Sprint Review at the end of each Sprint where the team presents the deliverables to the stakeholders.
Members of the meeting collaborate on what was done and what to do next.
The Product owner collects feedback and updates the Product Backlog (if needed) for future sprints.

\emph{Sprint retrospective:} The Sprint retrospective is a meeting, which held after the Sprint review and before the next Sprint Planning Meeting. 
Team members give feedbacks on what went well and not so well during the Sprint.
The team collaborates in identifying some improvements for future Sprints.

\begin{figure*}[htbp]
\centering
\includegraphics[height=11cm]{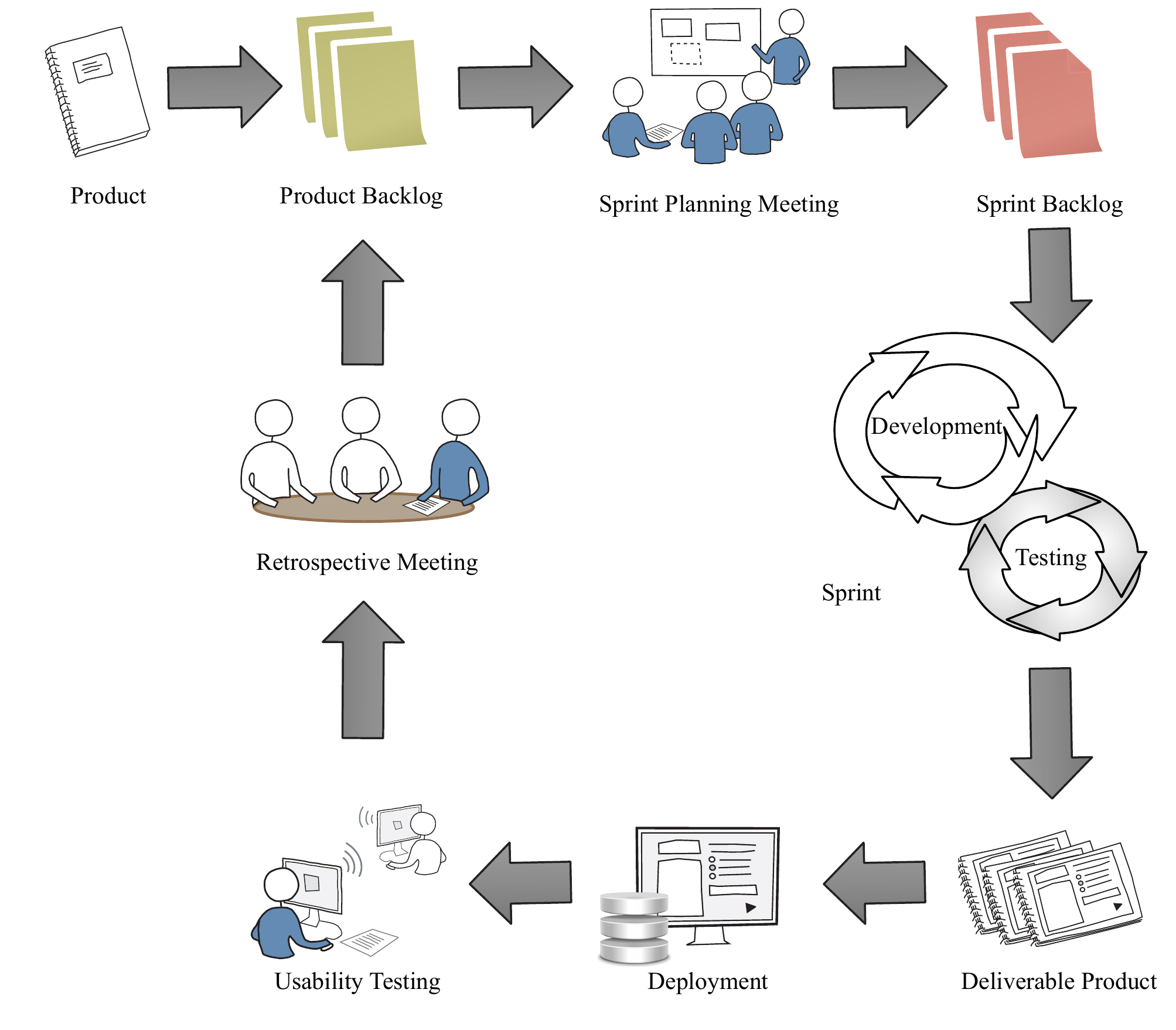}
\caption{The development of a Product using the Scrum process.}
\label{fig:scrum_flow}
\end{figure*}

Figure~\ref{fig:scrum_flow} shows how the development of a software product is done using a Scrum process.
This workflow incorporates Scrum Artifacts, Roles and Events (usability testing in the figure is not done by a Scrum team, but we incorporate it to show the full workflow of the product development).

\section{Product Backlog Rating (PBR)}
\label{PBR}

The Product Backlog (PBL) is the heart of a Scrum process. 
PBL is basically a prioritized list of requirements, or stories, or features, or things that  customers want, described using the customers' terminology~\cite{kniberg2007SXT}. 
We propose Product Backlog Rating (PBR) metric to measure the quality of the testing process in a Scrum process.
Note that PBR does not measure the quality of the end product, rather it measures how a testing process is performing in the long run.
PBR measures the quality of the Software Quality Assurance Layer.

PBR works with the PBL items. 
It assesses the test quality for each PBL item and directs the assessments to the evaluation of the Sprint test quality,  and eventually it presents the quality of the testing process in a Scrum team at a longer interval. 
Two types of quantifications are made for each PBL  item based on relevant factors. 

\begin{itemize}

\item  PBL Complexity Level (PCL): PCL gives a quantification of the testing hazards of a PBL.
It is based on some factors such as the complexity of the business logic of a PBL item, test data required  to test the PBL item, test estimation required to test the PBL item, a PBL item's internal and external dependencies, complexity of the test execution of the PBL item, business complexity at the GUI of the PBL item, third-party external interface integration hazards and data migration complexity.
Considering all the factors, PCL gives an indication of testing hazards of a PBL item. 

\item  Test Assessment Rating (TAR): TAR assesses how successful the testing process is in terms of finding bugs, uncovering severe bugs and lessening the number of bugs out of the test cases. 
TAR also considers a bug fixing ripple impact, requirement changes and the QA engineer's test confidence.
\end{itemize}

Let  we have $N$ number of PBL items $PBL_1,PBL_2,PBL_3, \dots ,PBL_N$. \\
PCL and TAR factors numbers are $P$ and $T$ respectively. \\
PCL factors are $(FP_1,FP_2,FP_3, \dots ,FP_P)\in FP$.\\
TAR factors are  $(FT_1,FT_2,FT_3, \dots ,FT_T)\in FT$.\\
PCL and TAR of the $k^{th}$ PBL item is based on the following equations.\\

\begin{equation}
PCL_{k}=  \frac{\sum_{i=1}^{ i \in P}FP_{i}*W_{i}}{\sum_{i=1}^{ i \in P}W_{i}}
\end{equation}

Where $W_i$ is the weight of the $i^{th}$ factor $FP_i \in FP$, $0< W_i \leq 1$. 
Weight represents the relative importance of the factors. 
Factor \emph{A} may have higher value than factor \emph{B} in terms of testing but Factor  \emph{B} may have higher importance due to the business need or other reasons. 
Weight $W$ balances this scenario by assigning higher value for factor B. 

\begin{equation}
TAR_{k}= \sum_{i=1}^{ i \in T} \frac{FT_{i}}{T}
\end{equation}

Using PCL and TAR of the PBL items, PBR can be calculated according to the following equation.

\begin{equation}
PBR=  \frac{\sum_{i=1}^{ i \in N}PCL_i*TAR_i}{\sum_{i=1}^{ i \in N} PCL_{i}}
\end{equation}


\section{Product Complexity Level (PCL) Factors}
\label{pcl}

\subsection{Complexity of Business Logic of the PBL Item}
The Product Backlog is an ordered list of ``what a sub-system will do''.
It consists of  requirements, bug fixes, non-functional requirements, etc.~\cite{Paetsch2003PBL}.
Although the Product Backlog items supplied by the Product Owner are further split into Sprint Backlog items by the Scrum team, we consider Product Backlog items as granular testable entities.  
For example, consider a web-based table in a software system that takes user input  from a text box and shows personalized output in the table querying remote database server.
This whole story could be incorporated in a Product Backlog item.
But a Scrum team will divide the Product Backlog item into multiple Sprint Backlog items, for example,  developing the GUI, processing user input, fetching data using query, collecting data using data adapter, connecting the data adapter with output module and more or similar technical Sprint Backlog items.
Unfortunately, from a QA engineering's perspective, not all the Sprint Backlog items are testable (e.g., query, data adapter) from the ``black box'' level. 
Even if we only consider the testable Sprint Backlog items, a myriad of  Sprint Backlog items~\cite{Gannon2013Scrum} might jeopardize the testing process in a sense that a QA engineer has to write and manage an astonishing amount of test cases (manual and automated).

Determining the business logic complexity of a PBL item is a challenging task.
Martin Fowler states that this is partly due to the business logic rarely fit any logical pattern; after all it is written by business people to capture business and an odd  variations can make a huge difference~\cite{Fowler2002Patterns}. 
Moreover, complexity assessment demands product related experiences and a well-established domain knowledge.
However, the iterative nature of the Scrum process could be leveraged.
During the rudimentary iterations, a QA engineer might struggle to comprehend the complexity, but later when some iterations will be delivered, he might be well-informed about the complexity (note Boehm's maxim  related to requirement understanding--- ``You don't know the requirements until the project is done''~\cite{BoehmMaxim}). 

We propose a simple, logical and validation based investigation inspired by the fact that  software is a formal description, hence logic provides  a vehicle for performing analysis of its behavior~\cite{Nuseibeh2000logic}.
A QA engineer could go extensively to the requirements and count validations and if-else conditions (or similar types of logics) in the requirement to assess the business complexity.
Creating a flowchart from the requirement and inferring logical conditions might help as well.

\subsection{Necessity of Test Data}

Test data are  the inputs given to a software program. 
The execution of a software module depends on the test data.
In software testing, two types of test data are prepared depending on whether positive or negative tests will be performed.
Test data for positive testing are prepared to ensure that the software module works as expected.
On the other hand, test data for negative testing are prepared to assess how the software module will work on atypical, exceptional, or unexpected input.
Naively designed test data might not allow to test all execution scenarios of a software module, hence could severely compromise the quality of the end product.
Researchers have already characterized test data and proposed techniques to generate test data~\cite{Zaeem2012testdata,samuel2010testdata,alakeel2010testdata,rizwan2012testdata}.
 
However, test data preparation is always challenging and complexity of the test data generation might contribute more hazards in testing.
For example, consider a PBL item which uploads Excel documents ---a user inserts a bulk amount of data in a pre-configured Excel template and uploads that in the system.
To test the PBL, the QA engineer requires to prepare an astonishing amount of test Excel documents (considering all positive and negative scenarios possible).
This task is also complex enough that an automatic generation demands critical thinking and significant efforts.
So, characteristics and volume of test data could reflect the testing complexity of the PBL.
Following points could be taken into consideration while assessing this factor.

\begin{itemize}
\item {Volume of the test data.}
\item{Complexity of the test data.}
\item{Modules required to configure for the test data, test environment configurations, and pre-steps needed.}
\item{Manual or automatic generation nature of the test data.}
\item{Reusability of the test data.}
\end{itemize}

\subsection{Test Estimation}
Test estimation is an estimation of the testing size, testing effort, testing costs and testing schedule for a software module under test in a specified environment using defined methods, tools and techniques~\cite{chemuturi2006test}.
The intuition behind including test estimation as a factor of determining a PBL item's complexity is that a high estimated test prone to reveal more faults. 
In~\cite{chemuturi2006test}, Chemuturi described following techniques to assess test estimate.

\begin{itemize}
\item Delphi technique 
\item Analogy-based estimation 
\item Software size based estimation 
\item Test case enumeration based estimation 
\item Task (Activity)-based estimation 
\item Testing size-based estimation 
\end{itemize}

We propose to use any of the techniques (based on the QA engineer's expertise) and estimate test in man-days. 
Later, we can give a rating to this factor based on the man-days. 
Factor rating will be high for those PBL items those demand high test estimations.

\subsection{PBL Item Inter-dependency}

Generally speaking, PBL items' inter-dependencies could make a simple task difficult; not only in terms of development, but also in terms of quality assurance.
Consider two tasks $task_A$ and $task_B$ of a PBL item.
There are four possible scenarios in terms of dependencies, all of them are listed below.

\begin{itemize}
\item  $task_A$ and $task_B$ are independent.
\item  $task_A$ is done and $task_B$'s dependency on $task_A$ affects  $task_B$'s development, and vice versa.
\item  $task_A$ is done, and $task_B$ is also done, but development of $task_B$  after $task_A$ affects $task_A$, and vice versa.
\item  $task_A$ is done, and $task_B$ is also done, but their developments affect (bring ripple impact) to another task.
\end{itemize}

Furthermore, consider a real-world example of a PBL item, which has four tasks.
For example, a simple list page with add, edit and delete functionalities. 
Now add, edit, delete tasks have an impact on the list page. 
Again deleting an item from a list page can trigger a concurrent issue of editing at the same time. 
These are some examples of simple PBL items' inter-dependencies. 
So, we argue that PBL items' inter-dependencies could contribute to the complexity of the PBL item testing.
Depending on inter-dependencies, we can give a high rating for a highly inter-dependent PBL  item and a low value for a less inter-dependent PBL item.

\subsection{PBL Item  External Dependency}
It is generally accepted that most of the PBL items in a software system are not fully independent.
The extraction and exploitation of dependencies has been a subject of research for a long time~\cite{Sangal2005UDM,gall1998coupling,cataldo2009software}.
In a highly coupled system, testing hazards are prevalent due to PBL items' external dependencies.

Consider two PBL items: $PBL_A$ and $PBL_B$.
There could be  four possible scenarios in terms of PBL items' dependencies, similar to PBL items' inter-dependencies, all of them are listed below.

\begin{itemize}
\item  $PBL_A$ and $PBL_B$  are independent.
\item   $PBL_A$  is done and $PBL_B$ dependency on $PBL_A$ affects  $PBL_B$'s development, and vice versa.
\item  $PBL_A$ is done, and $PBL_B$ is also done, but development of $PBL_B$  after $PBL_A$ affects $PBL_A$, and vice versa.
\item  $PBL_A$ is done, and $PBL_B$ is also done, but their developments affect (bring ripple impact) to another PBL.
\end{itemize}

Assuring quality  of the PBL items  which have external dependencies are challenging for a number of reasons.
First, it is difficult to estimate when testing is done for those PBL items.
Depending on the scenarios we listed above, a well tested PBL item might require a full test phase to run after development of another PBL.

Second, bug propagation is high among dependent PBL items.

Third, it is  difficult to generate test cases for a PBL item which is highly dependent on others.
Because, not only the developed features, but also the test cases are also highly dependent.
So, depending on the external dependencies, the QA engineer can give a high rating for highly external-dependent PBL item and a low value for a less external-dependent PBL item.
\subsection{Test Execution Complexity}
A PBL item might involve less business logic, it might have less internal or external dependencies, or it might require less test efforts, but still test execution complexity could make quality assurance tasks difficult.

We present some real-world examples below.
\begin{itemize}
\item  Concurrency testing has a  high level of complexity than a single data update testing.
\item  Real time systems have deadlines to meet, with consequences for failure.
\item  A briefing, production workflow in a content management system might produce an output in a PDF format which associates a precise checking of semantic mapping.
\item  Complicated output generation takes time in a system which adds some delay overhead.
\item  Complex log analysis.
\item High number of granular steps to reach the test result.
\item Time consuming post test cleanup.
\end{itemize}

\subsection{Logic on Graphical User Interface (GUI)}
In the age of Web $2.0$, software systems are becoming more interactive, allowing users to add their own data in the system~\cite{orily2007web}.
These systems are providing more interfacing equipments, for example, sophisticated graphical user interface (GUI).
As a consequence, the GUI has become more complex and data-driven.
As the businesses have to accommodate interactive user behaviors, many business requirements are reflecting on the GUI.
Business requirements at GUI implies more validations, and hence testing.
Moreover, testing those complex GUIs on multiple browsers and on multiple platforms is challenging. 

\subsection{Bug Assumptions}
Bug assumptions could be leveraged to determine testing hazards of those PBL items that have historical bug records.
Ideally, in any type of software development it is difficult to foresee bugs (although researchers tried to predict bugs, e.g., using bug databases~\cite{Zimmermann2008Bugs}).
But in Scrum, some features' developments are incremental in a sense that several iterations are dedicated to enhance those features.
So, PBL items associated with theses features contain historical bug records.
Historical bug records might help assuming future bugs, at least qualitatively.
And it is commonly accepted that bug prone features require significant testing efforts~\cite{kaner2004software}.  

\subsection{Data Migration Complexity}
Data migration involves extracting data from a source system, correcting existing bugs (if needed), reformatting, restructuring and loading the data into a replacement target system~\cite{pick2001data}.
Simply, a migration process is a set of mapping rules and transformation functions.
One of the motivations behind the Scrum process is to accommodate requirement changes from the stakeholders.
So, what happens in the process is that a working software is delivered to the end users and stakeholders, and they evaluate what more the system should do.
This leads to data migration activities;  new features are added (sometimes replacing the old) and user generated data are migrated to serve new features.
However, following requirements related to data migration make a PBL item testing complex.

\begin{itemize}
\item The new data model supports some new application features.
\item The new data model no longer supports some old  application features.
\item A gray box testing (i.e., database) is not enough; GUI-based testing might be required. 
\item Migration might involve cascade---migration might happen in multiple steps where each migration step initiates another migratory step.
\item Some user generated data might be bad, they might not be transformed.

\end{itemize}

\subsection{$3^{rd}$ Party Support / External Interface Integration}
Some PBL items involve integrating $3^{rd}$ party supports or external interfaces.
For example, a PDF output generation of a PBL  item might feed input into a $3^{rd}$ party PDF generator engine.
$3^{rd}$ party supports are always considered a black-box; the only thing the team knows about the black-box is what the black-box takes (input) and what it produces (output).
Thus, a quality assurance work of those PBL items begins from the $3^{rd}$ party supports---an extensive testing is required.  
This might require a quality person to learn new technologies, gather domain-specific knowledge involved with those $3^{rd}$ party supports.

\subsection{Context Driven Factors}
Software projects unfold over time in ways that are often not predictable by the Scrum team.
As such, best practices in one context become obsolete in another context.
Researchers argue that a context-driven testing approach~\cite{fiedler2009putting,kaner2008lessons} might be appropriate for Agile software development.
In context-driven testing, rather than trying to apply ``best practices'', testers accept that very different practices that will work best under different contexts~\cite{kaner2008lessons}.
So, depending on the context of a PBL item, context driven factors should be considered in defining testing hazards of a PBL item. 

\section{Test Assessment Rating (TAR) Factors}
\label{tar}

\subsection{Bugs Count}
IEEE Standard 610 (1990) defines a test case as ``\emph{A set of test inputs, execution conditions, and expected results developed for a 
particular objective, such as to exercise a particular program path or to verify compliance with a specific requirement}"~\cite{moore1998software}. 
In~\cite{Kaner2003testcase} Cem Kaner describes several objectives of a test case which includes finding bugs, maximizing bug count, assessing conformance to specification, conforming to regulations, finding safe scenarios for using of the product, and assessing quality.
In Scrum, we want to point out that test cases help a QA engineer to find out problems in the requirements or in the design of the software system since the QA engineer is involved throughout the Sprint, contrary to traditional development methods where testing starts after all developments are done.
As such, a QA engineer can write test cases possessing a distant vision in Scrum.
Although exploratory testing is a common approach in Scrum, we argue that finding bugs using test cases quantifies test case quality. 
The more bugs are found using pre-defined test cases, the more it guarantees that exhaustive testing of the software is done and this factor rating increase.

\subsection{Severity of the Bugs} 
Severity is a fundamental measure of the impact of a bug. 
All bugs are not equal in terms of severity; some are trivial but some are severe.
For example, independent bugs can be directly detected and removed, but mutually dependent bugs can be removed if and only if the leading bugs have been removed~\cite{Huang2006Fault}.
This means that the leading bugs have severe impact than the bugs that follow them.
So, a bug count does not quantify the testing quality if the bugs are minors. 
Severity of the bugs should be considered in counting bugs.

\subsection{Bug Fixing Ripple Impact} 
A common scenario in software testing is that when a bug is fixed, it breaks other components.
Sometimes bug fixing causes ripple impact---a small fix effects a series of components.
Also experiences show that this ripple impact hampers the quality of the product. 
So, we propose to consider the ripple impact while assessing the testing.
To rate the ripple impact we can measure how much percentage of the tests we have to do again.
If too much ripple impacts and testing efforts are required, then this factor rating will go down.

\subsection{Number of Bugs Missed by the Test Case}
The test case is the main engine of the testing process. 
Ideally, it should cover all sorts of test scenarios, possibly one could imagine.
However, it is also very natural that some bugs will be revealed on some test scenarios which are not covered in the test case. 
Test cases are prepared to give a complete thought of the operationality of the application.
We consider too much and too severe bugs missed by the test case as a sign of lack of understanding of the operation of the software.
Hence, those missing bugs could be taken into consideration while assessing the testing process. 

\subsection{Requirement Changes}
Research~\cite{stark1999examination} shows that requirement changes impacts the cost, schedule, and quality of the resulting product.
In fact, Gorschek and Davis present~\cite{gorschek2008requirements} a conceptual framework for assessing the impact of requirement changes.
In Scrum, if requirements are changed or added in the middle of the Sprint then we cannot expect good quality of a PBL item relative to a PBL item without any requirement changes. 
So, we propose to consider requirement changes while assessing the testing based on the intuition that too much or too late requirement changes affect the testing process negatively. 

\subsection{Test Confidence}
Test confidence is always subjective.
It varies from QA engineers to QA engineers and projects to projects.
However, it is also true that an experience  QA engineer can always make some comments on the testing (e.g., whether he is confident enough that enough testing has been done and the product is close to bug-free) based on his experience and intuition.
Test confidence factor values a QA engineer's intuition and experience.
However,  organizations are free to exclude this factor if they do not want to include a factor that is  too much subjective.

\section{A case study from SoftwarePeople}
\label{casestudy}

We present a  case study taken from SoftwarePeople.
SoftwarePeople is a Denmark-based Marketing Operation Management (MOM) solutions provider for large international clients such as Dell, Intersport, MediaSyd etc.
The MOM solutions are briefing, production workflow and content management systems, which enable the clients and their vendors to optimize the production of marcom material. 
They allow the briefing of any marketing project, from a simple banner to a complex website or catalog. 
SoftwarePeople has been practicing Scrum for about eight years and PBR enables the organization to get a comprehensive view of the test quality over several Sprints. 
SoftwarePeople has 6-7 developers and a QA engineer in each Scrum team and its typical length of a Sprint is 3-4 weeks. 
The following PBL items have been taken from a three weeks long Sprint.
We describe the functionality of the PBL items, their internal and external relations with other modules, and hazards that might happen in testing. 
This will help readers to understand the ratings that a QA engineer has assigned for PCL and TAR factors.

\begin{itemize}
\item \textbf{\emph{PBL item \#1: Enhanced Text Bank functionality to enable predefined formatted Microsoft excel documents to configure the Text Bank.}}

Text Bank is an existing GUI-based text management system, used by users from multiple continents.
Text Bank maintains the translation of a text for multiple countries.
The functionalities of the Text Bank include the addition of a text (with or without country-specific translation), edition and deletion of a text, basic and advanced search functions etc.
End users use the Text Bank to produce briefing materials for an array of counties---for example, consider a laptop or computer catalog of a client, which will be distributed in multiple countries using country-specific translations.
In some cases, users upload a bulk amount of texts and translations using Excel documents.
The new requirement asks an enhancement, which will enable the users to export texts from the Text Bank in a Spreadsheet and enable users to import the  Spreadsheet with modifications made by them.

\item \textbf{\emph{PBL item \#2: Color scheme for Activity module: Enable users to view user configurable status color in activity searching.}}

An Activity is a core of a MOM solution.
Activities are used to produce different types of output such as product catalog, banner.
An Activity consists of several modules.
A metadata module consists of basic description of the activity such as the type of the activity, the country where the activity will be exposed, the exposure date, the briefing date etc.
The asset module consists of different assets related to the activity such as texts and images. 
There is also a workflow module---using this module users can manage the workflow of an Activity. 
Each Activity has some statuses; users are assigned to those  statuses from the workflow module.
Users can search the activities based on the metadata and activity statuses.
This PBL item asks for user configurable/ defined color scheme so that after searching an activity, a user can see his defined color according to the status of the activity in Activity search list page.

\item \textbf{\emph{PBL item \#3: Time Zone and Date: time and date fields should be according to the users' time zone settings.}}

An Activity involves a workflow management, participated by people from multiple time-zones.
At the time of this work, Activity time and date format were based on the server settings (where the Activity was stored).
This new requirement requires storing the time and date format according to user settings. 
The requirement requires  a Time and Zone tab in an existing ``My Settings'' page of the user.
Users will be able to change the Time Zone from a list of all available time zones.
Moreover, there will be date formatting options available for users, for example a combination of day, month, year and AM:PM, $24$H.
All time and date fields of the system will be presented to a user according to his settings.

\item \textbf{\emph{PBL item \#4: Incorporating ``Theme'' and ``Campaign'' in the Text Bank and enabling text search according to ``Theme'' and ``Campaign''.}}

This PBL item is about adding two extra fields for each text in the Text Bank. 
Search functionality requires to be modified so that search is possible with Theme and Campaign.

\item \textbf{\emph{PBL item \#5: Change country functionality for Activity: Current stick to the behavior of the Activity needs to be changed so that country of an Activity can be changed.} }

An Activity is associated with a country, which is not editable.
However, the new requirement asks to provide an option so that users can change the country of an Activity.
This requirement triggers some cascade of changes.
Each Activity is associated with some assets such as texts and images. 
But those assets are country-specific and have distinguishable properties for countries (e.g., text translation for countries).
Activity country change will hold those assets if they are valid for the changed country, otherwise they will be discarded. 
Moreover, in the workflow module of an Activity, for each Activity status there exists a set of approval groups.
Changing the country of the Activity might require to alter the  approval groups also.
So, it appears that a simple change in the Activity metadata creates a lot of changes in the underlying business logic.
\end{itemize}

We use following abbreviations for PCL factors.
\begin{itemize}
\item CBLBP- Complexity of Business Logic of the PBL
\item NTD- Necessity of Test Data
\item TE- Test Estimation
\item PID- PBL item Inter-dependency
\item PED- PBL item External-dependency
\item TEC- Test Execution Complexity
\item LGUI- Logic on Graphical User Interface (GUI)
\item BA- Bug Assumptions
\end{itemize}

We use following abbreviations for TAR factors.
\begin{itemize}
\item BC- Bugs Count
\item SB- Severity of the Bugs
\item BFRI- Bug Fixing Ripple Impact
\item NBMTC- Number of Bugs Missed by the Test case
\item RC- Requirement Changes
\item TC- Test Confidence
\end{itemize}

\begin{table*} [ht]
\centering
\begin{tabular}{ | l| l| l| l| l| l| l| l| l| l| l |  }
  \hline
  PBL& &\multicolumn{8}{|c|}{PCL factor rating ($1$ to $5$)} & PCL \\
\cline{3-10}
    &  & CBLBP  & NTD  & TE  & PID & PED  & TEC & LGUI  & BA  &\\
    \hline
   PBL1& Factor Value & 5	 &5	&5	&4.5	 &5	&5	&5	&5	&4.93 \\
   \cline{2-10}
   & Weight &1	&1	&1	&1	&1	&1	&0.4	&1& \\
   \cline{2-10}
   &Factor Rating& 5	&5	&5	&4.5		&5	&5	&2	&5& \\
   \hline
   
      PBL2& Factor Value  &3	&4	&3	&3.5  &4	&3	&4	&2	&3.39\\
   \cline{2-10}
   & Weight &0.7	&1	&0.6	  &0.8	&0.7	  &0.5	&0.7	 &0.6& \\
   \cline{2-10}
   &Factor Rating& 2.1	&4	&1.8	  &2.8	&2.8	 &1.5	 &2.8	  &1.2& \\
\hline

 PBL3& Factor Value & 3	 &3	&4	&3	&3	&4	&5	&2	&3.62\\
   \cline{2-10}
   & Weight &0.7	&0.7	 &1	&0.5	  &0.5	&1	&1	&0.49& \\
   \cline{2-10}
   &Factor Rating& 2.1	&2.1  &4	&1.5	  &1.5	&4	&5	&0.8& \\
\hline

PBL4& Factor Value & 2	&2.5	 &2.5  	&2	&3	&2.5	 &2.5 	&2	&2.42\\
   \cline{2-10}
   & Weight &0.6	&0.7	 &0.8  	&0.6	  &0.8	&0.8  	&0.6	 &0.5& \\
   \cline{2-10}
   &Factor Rating& 1.2	&1.8	 &2	&1.2 	 &2.4 	&2	&1.5	& 1& \\
   
 \hline 
   
PBL5& Factor Value &5	&4	&5	&4	&3	&5	&3	&5	&4.37\\
   \cline{2-10}
   & Weight &1	&1	&0.7	 &0.8 	&0.5	 &1	&0.7	 &1& \\
   \cline{2-10}
   &Factor Rating& 5	 &4	&3.5	 &3.2  &1.5	&5	&2.1	&5& \\
\hline
   
  \hline
\end{tabular}
\caption{PCL factors (with empirical values)}
\label{pcl_sp}
\end{table*}

\begin{table*} [htbp]
\centering
\begin{tabular}{ | l| l| l| l| l| l| l| l| l |   }
  \hline
  PBL& &\multicolumn{6}{|c|}{TAR factor rating ($1$ to $5$)} & TAR \\
\cline{3-8}
    &  & BC  & SB  & BFRI  & NBMTC & RC  & TC & \\
    \hline
   PBL1& Factor Rating &5	&5	&2	&3	&2	&3.5  	&3.41 \\
    \hline
   PBL2& Factor Rating &4	&2	&5	&5	&5	&4.5		&4.25 \\
   \hline
    PBL3& Factor Rating &5	&3	&5	&5	&5	&4.5		&4.58 \\
   \hline
     PBL4& Factor Rating &4	&2.5	 &4	&3	&5	&3.75	&3.70 \\
    \hline
      PBL5& Factor Rating &2	&5	&3	&2	&1	&2.5		&2.58 \\
    \hline

    \hline
\end{tabular}
\caption{TAR factors (with empirical values)}
\label{tar_sp}
\end{table*}

Table~\ref{pcl_sp} and Table~\ref{tar_sp} shows empirical values (all are out of $5.00$) of PCL and TAR factors for the Sprint, assigned by a QA engineer.
The QA engineer assigns the PCL factors after the task breakdown.
During the Sprint he might change the PCL factors based on changing requirements.
However, TAR factors are assigned after the testing is done for the Sprint.
From the PCL and TAR of the PBL items, we can calculate PBR score of the Sprint as follows.

\begin{align*}
PBR
	&	=  \frac{\sum_{i=1}^{ i \in N}PCL_i*TAR_i}{\sum_{i=1}^{ i \in N} PCL_{i}} \\
	&	= 3.11
\end{align*}

SoftwarePeople has defined PBR score interpretations. 
According to  the interpretations, test quality of this Sprint earns a moderate level.
We use the interpretations shown in Table~\ref{interpretation}.
These interpretations may vary organizations to organizations, our interpretations are based on historical test assessments.

\begin{table} [ht]
\centering
\begin{tabular}{ | l| l| }
  \hline
  PBR score & Interpretation\\
  \hline
  $5$ & Excellent\\
  \hline
  $4$ & Good\\
    \hline
  $3$ & Moderate\\
    \hline
  $2$ & Bad\\
    \hline
  $1$ & Worst\\
    \hline
\end{tabular}
\caption{Interpretations of PBR scores.}
\label{interpretation}
\end{table}

\begin{figure}[htbp]
\centering
\includegraphics[height=5.5cm]{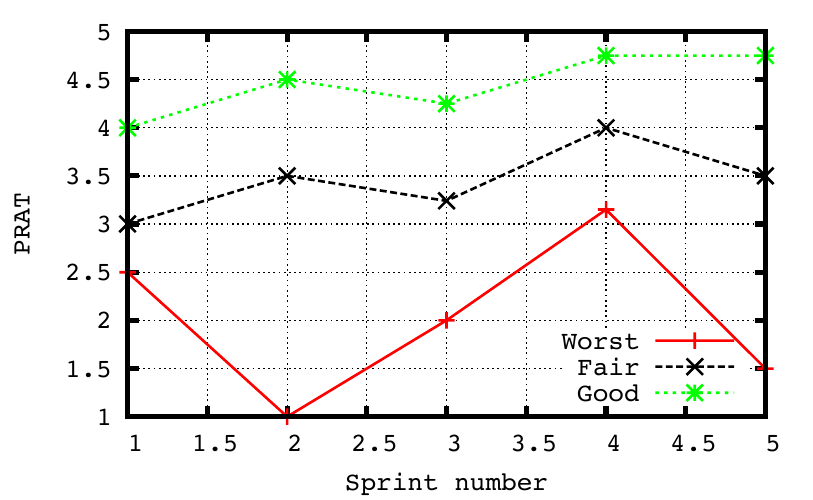}
\caption{PBR plot against Sprints }
\label{fig:PBR_samle}
\end{figure}

A PBR plot over consecutive Sprints can provide a comprehensive overview of the testing process in a Scrum team.
We present a sample plot and interpretations  in Figure~\ref{fig:PBR_samle}.

\section{Related Work}
\label{related}
Software metrics quantify specific attributes of a software product or the process of the software development~\cite{Grady1994Metrics}.
A wide body of research attempts to numerically evaluate the  quality characteristics of a software product.
Also, software process improvement has been studied for a long time.
However, to the best our knowledge, we are the first to evaluate the quality of the software testing in an Agile process.

To address the product quality characteristics and subcharacteristics, the Joint Technical Committee $1$ of the International Organization for Standardization and International Electrotechnical Commission defined a set of software product quality standards known as ISO/IEC 9126~\cite{jung2004measuring} (see a complete listing in Table~\ref{ISO}).
They defined six high-level product quality characteristics which are as follows.
Functionality--extent to which each function of the software system operates in conformance with the requirement specification; reliability---extent to which a software can be expected to perform its intended function with required precision; usability---effort required to learn, operate, prepare input, and interpret output of a program; efficiency---the amount of computing resources and code required by a software to perform a function; maintainability---effort required to diagnose and fix a bug in an operational software and portability---effort required to transfer a software from one hardware configuration and/or software system environment to another~\cite{cavano1978framework}. 
These quality characteristics have been studied and researchers proposed a wide range of metrics to measure them (e.g.,~\cite{bertoa2006measuring,nagappan2003towards,goel1985software,oman1992metrics,coleman1994using,washizaki2003metrics,henry1981software}).

\begin{table} [ht]
\centering
\begin{tabular}{ | l| l| }
\hline
   Characteristics & Subcharacteristics\\
  \hline 
  \hline
  Functionality & Suitability, accuracy, interoperability,\\
&security, functionality compliance\\
   \hline
  Reliability & Maturity, fault tolerance, recoverability, \\
  &reliability compliance\\
    \hline
  Usability & Understandability, learnability, operability, \\
  &attractiveness, usability compliance\\
    \hline
  Efficiency & Time behavior, resource utilization, efficiency compliance\\
    \hline
  Maintainability & Analyzability, changeability, stability,\\
   &testability, maintainability, compliance\\
\hline
  Portability & Adaptability, installability, replaceability, \\
  & coexistence, portability compliance\\
  
    \hline
\end{tabular}
\caption{Characteristics and subcharacteristics in ISO/IEC $9126$.}
\label{ISO}
\end{table}


Software process improvement attempts to make the software process more efficient and ensure end product quality by continuous assessment and adjustment of the process~\cite{petersen2010software}.
Software process improvement research is motivated by the intuition that  process quality and process quality are inter-related~\cite{
cugola1998software,fuggetta2000software,kitchenham1996software}.
The Capability Maturity Model (CMM)~\cite{paulk1995capability} is an initial attempt to increase software development capability and process maturity~\cite{paulk1993comparing}.
This is a framework that presents prime elements of an effective software process. 
The model describes an evolutionary improvement path for a software development process from an ad hoc, immature process to a mature, disciplined process, in a path laid out in five levels~\cite{paulk1995capability}. 
Similarly, the Capability Maturity Model Integration (CMMI)~\cite{ahern2004cmmi,team2002capability} and ISO/IEC $15504$~\cite{emam1997spice,jung2001relationship} propose various measurements. 
The CMMI model provides measurement recommendations for each process area as an informative supplement to the required components of the model. 
The ISO/IEC $15504$ defines a process measurement framework and suggests that the process improvement has to be confirmed. 
Quality Improvement Paradigm (QIP)~\cite{basili1994software} is based on the principle that software discipline is evolutionary and experimental and all project environments and products are different. 
The QIP cycle is comprised of two closed loop cycles---organizational and project  cycle.
Organizational feedback cycle gives feedback to the organization after the completion of the project.
Project cycle provides feedback to the project during the execution phase in order to prevent and solve problems, monitor and support the project~\cite{basili1994software}.
McFeeley proposed IDEAL~\cite{mcfeeley1996ideal}, a five phase process improvement model, which provides a continuous loop
through several steps.

Recently, traditional software development methodologies have been challenged by Agile software development methods such as Scrum.
Agile processes emphasize on context-specific adaptive development, workable product, customer satisfaction and small iterative development.
In~\cite{salo2007iterative} Salo and Abrahamsson points out underlying differences of traditional and Agile software development from the viewpoint of process improvement and conclude that new mechanisms are needed to fit the context of the Agile software development.
The main distinguishing points they mention are (1) a traditional software development process control is maintained from organization level where self-organizing teams manage the process control in an Agile development; (2) primary means of knowledge transfer in traditional software development is document based but in Agile it is  face-to-face communication; (3) a traditional software development process focuses on improvement of organizational software development processes or future projects, but in Agile development the focus is on the improvement of daily working practices of ongoing project. 
Cockburn and Highsmith describe ``The People Factor'' in~\cite{Cockbur2001Agile,cockburn2003agile}. 
They argue that Agile development will be effective if (1) it can reduce the cost of moving information between people, and (2) reduce the elapsed time between making a decision to seeing the consequences of that decision. 
For the first scenario, they suggest to place people physically closer, replace documents with talking in person and at whiteboards, and improve the teamÕs amicability. 
For the second scenario, they emphasize on making user experts available to the team or, even better, part of the team and working incrementally~\cite{Cockbur2001Agile}.
Our work is related to an Agile process (Scrum) improvement.
However, we have focused on the process of testing in the Scrum process.
We have proposed a metric PBR that evaluates the testing process in Scrum and showed the metric ``in action'' in an industry setting.
The metric could be used to monitor the testing process in the Scrum.
However, at the same time it is the responsibility of a Scrum team to find out what testing strategies might improve the metric.

\section{Summary}
\label{conclusion}

In this paper, we have proposed a metric to evaluate the testing process in Scrum. 
Quality is a complex and subjective concept, thus hard to quantify.
We attempt to offer a metric that gives a numerical score to the quality of the testing process in Scrum.
We have discussed the Scrum process first, emphasizing testing in Scrum.
We have leveraged SoftwarePeople's successful Scrum implementation experience to describe the Scrum process.
Our proposed metric PBR is based on the complexity of the Product Backlog and test assessment rating.
The metric can provide an additional framework for quality management. 
We provide a case study from SoftwarePeople to show how the metric is used in practice.

The quality of a testing process cannot be precisely measured as a lot of development-driven factors are involved.
These factors can alter the quality of the product. 
Not only development-driven factors, but also deployment-driven factors could jeopardize the quality of the product.
For example, a bug in the automated deployment scripts could end up not deploying some modules of a product.
In this case, all functionalities dependent on those modules will malfunction. 
Our hope is to assess the test quality up to some degrees.

Quality management in Scrum is a collaborative approach.
In traditional software development methods, QA engineers are solely responsible for ensuring product quality.
However, Scrum suggests developers  conduct tests like Unit testing~\cite{watkins2009agile} or TDD~\cite{stamelos2007agile} before delivering the features to a QA engineer.
In this work, we have only considered QA engineers' efforts in determining test assessment ratings.
In future, we will attempt to include developers' testing contributions in assessing the test process.

\bibliographystyle{splncs} 
\bibliography{Bibtex}








\end{document}